\documentclass[preprint,12pt]{elsarticle}
\usepackage{epsf}
\usepackage{graphicx}
\usepackage{amsmath}
\journal{Physica D}
\usepackage{float}
\begin{document}
\begin{frontmatter}

  \title {Quantization of $\beta$-Fermi-Pasta-Ulam Lattice with Nearest and Next-nearest Neighbour Interactions}

\author[1]{Aniruddha Kibey}
\author[1]{Rupali Sonone}
\author[1]{Bishwajyoti Dey\footnote{bdey@physics.unipune.ac.in}}
\author[2]{J. Chris Eilbeck}
\address[1]{Department of Physics, University of Pune, Pune - 411007, India}
\address[2]{Department of Mathematics and Maxwell Institute,
  Heriot-Watt University, Riccarton, Edinburgh, EH14 4AS, UK}

\date{\today}

\begin{abstract}

  We quantize the $\beta$-Fermi-Pasta-Ulam (FPU) model with nearest and
  next-nearest neighbor interactions using a number conserving
  approximation and a numerically exact diagonalization method.  Our
  numerical mean field bi-phonon spectrum shows excellent agreement
  with the analytic mean field results of Ivi\'{c} and Tsironis ((2006) Physica
  D 216 200), except for the wave vector at the midpoint
  of the Brillouin zone.  We then relax the mean field approximation
  and calculate the eigenvalue spectrum of the full Hamiltonian.  We
  show the existence of multi-phonon bound states and analyze the
  properties of these states by varying the system parameters. From the
  calculation of the spatial correlation function we then show that
  these multi-phonon bound states are particle like states with finite
  spatial correlation. Accordingly we identify these multi-phonon
  bound states as the quantum equivalent of the breather solutions of
  the corresponding classical FPU model. The four-phonon
  spectrum of the system is then obtained and its properties are
  studied. We then generalize the study to an extended range
  interaction and consider the quantization of the
  $\beta$-FPU model with next-nearest-neighbor
  interactions.  We analyze the effect of the next-nearest-neighbor
  interactions on the eigenvalue spectrum and the correlation
  functions of the system.
  \let\thefootnote\relax\footnote{Accepted in Physica D}
\end{abstract}


\end{frontmatter}


\section{Introduction}
The Fermi-Pasta-Ulam (FPU) model marked the beginning of the field of
nonlinear dynamics.  It was initially introduced to examine the
possibility of equipartition of energy in a chain of coupled nonlinear
oscillators in the thermodynamic limit \cite{1}.  Since then this
model has been used to study many important problems in non-equilibrium
statistical mechanics, in particular heat conduction in a 1D chain
\cite{2}.  Other interesting aspects of the dynamics of FPU lattice are
the occurrence of intrinsic localized modes (ILMs) or discrete
breathers (DBs). ILMs or DBs are time periodic and spatially localized
solutions of the nonlinear lattice and occur due to the balance
between nonlinearity and discreteness.  ILMs have been experimentally
observed in various physical systems ranging from lattice vibrations
in crystals and magnetic solids, Josephson junction, photonic crystals
etc. ILMs have been used for modeling observed physical processes in
polymers and bio-polymers such as, folding in the polypeptide chain,
targeted breaking of chemical bonds, charge and energy transport in
conducting polymers etc., see \cite{3} for a review.  MacKay and Aubry
rigorously established the existence of breather excitations in
lattices under a wide range of conditions \cite{4}.  Exact compact
breather-like solutions of the two-dimensional FPU lattice were also
obtained \cite{5}.  The effect of long-range interactions, the
geometry of the chain and nonlinear dispersion on the dynamics of
breathers in two-dimensional FPU lattices, and the usefulness of these
solutions for energy localization and transport in various physical
systems are reported in \cite{6}.

It is well known that the integrable classical field equations, such
as the sine-Gordon equation \cite {7}, nonlinear Schrodinger equation
\cite{8} and modified and extended KdV equations \cite{9}, support
classical breather solutions which are a bound state of soliton and
antisoliton pair with continuum classical spectrum.  On the other hand
the Bohr-Sommerfeld semi-classical quantization for classical
breathers shows discrete excitations which are bound states of
small-amplitude wave excitations \cite{10}.  Faddeev and Korepin, while 
studying the quantum theory of solitons for integrable systems, showed that the breathers of classical integrable
field equations are indeed multi-phonon bound states \cite{11}.  These
studies led to the question of the nature of quantum breathers in
nonlinear lattices.  Even though there has been some studies along
this line for the nonlinear Klein-Gordon lattice and discrete
nonlinear Schrodinger equation \cite{12}, not much is known about the
nature of quantum breathers in FPU lattice model.
The problem of quantization of classical discrete breathers in FPU
model is difficult to solve exactly due to the nonlocal and anharmonic
nature of interactions in the model, which give rise to number non
conserving terms in the Hamiltonian.  Various truncation methods have
been used to obtain the approximate eigenspectra of the system. Ivi\'{c}
and Tsironis \cite{13} studied analytically the discrete breather
quantization problem in a $\beta$-FPU lattice by truncating the
Hamiltonian to include only number conserving terms, and analyzed
the bi-phonon sector of the model under the mean field approximation.
The number conserving approximation is equivalent to rotating wave
approximation used for studying the classical discrete breathers. Ivi\'{c}
and Tsironis showed the occurrence of bi-phonons of two types, the
on-site and nearest neighbor site and interpreted these bi-phonons as
the quantum counterparts of the classical FPU discrete breathers.
Xin-Guang and Yi \cite{14} showed the occurrence of a third type, a
mixed bi-phonon state, in a $\beta$-FPU model from their numerical
exact diagonalization study. However they  did not study the four-phonon eigenspectra, the
correlation functions, or the effect of the long range interactions on
the eigenspectra of the system.  Riseborough studied analytically the
quantized breather problem on FPU lattice in the two-phonon sector by
truncating the equation of motion of the two-phonon propagator, using
the ladder approximation in which the single-phonon self-energies are
neglected \cite{15}.  From the poles of the two-phonon propagator he 
obtained isolated frequencies which represent the discrete dispersion
relations for the collective breather excitations. Recently the bound
states of four bosons in the quantum $\beta$-FPU lattice have been reported
in \cite{16}.  However, they did not study the effect of the long
range interactions and the spatial correlation functions of the system.

In this paper we study this problem by numerically exact diagonalization
using the number state basis and including only number conserving
terms. The numerically exact diagonalization method has been
successfully applied to the Bose-Hubbard and modified Bose-Hubbard
Hamiltonians (\cite{17}-\cite{19}). The purposes of our
studies are manifold: (i) to check how the analytical results of Ivi\'{c}
and Tsironis \cite{13}, which are obtained under the mean field
approximation and in the large eigenvalue limit, compare with the
numerically exact diagonalization mean field results; (ii) to go beyond
the mean field approximation and obtain numerical eigenspectra of the
system; (iii) to study the problem for the higher quanta sector; (iv) to
calculate the spatial correlation functions to identify and establish
the main characteristics of the spatial localization of the discrete
breathers in the quantum excitation spectrum and (v) to generalize the
problem to examine the effect of increasing range of interactions, for
which we study the $\beta$-FPU model with next-nearest-neighbor (NNN)
interactions.  Interestingly, we find that the analytical mean field
results in the large eigenvalue limit of Ivi\'{c} and Tsironis \cite{13}
agree very well with our numerically exact diagonalization results.  We
also find that the eigenspectra for the complete Hamiltonian (beyond
the mean field case) is quite different as compared to that of the mean
field case. Contrary to the mean field case, the nearest neighbor
site bi-phonon mode do not get distinctly separated from the
quasi-continuum band.  This is in agreement with \cite{14}. The
four-phonon spectrum shows that the states which belong to the free
phonon band now split into several bands due to the effect of
nonlinearity. The isolated band corresponds to discrete
breathers. A similar splitting of bands for the higher quanta sector of the
Klein Gordon type nonlinear lattice has been reported by Wang et al.\
\cite{20} from their numerically exact diagonalization calculations
using an Einstein phonon basis.  The spatial correlation functions for
the on-site bi-phonon, as well as the four-phonon states, show strong
localization at particular sites.  This shows the particle like nature
of the states which identifies these states with corresponding
classical on-site discrete breathers. Similarly, the inclusion of the
NNN interaction in the $\beta$-FPU model changes the lattice
periodicity and its effect on the eigenspectra is clearly seen.

It may be noted that the number conserving approximation cannot be used
for $\alpha$-Fermi-Pasta-Ulam lattice. This is because the cubic
interaction terms in the  $\alpha$-Fermi-Pasta-Ulam Hamiltonian are all
number non-conserving.

It is appropriate to mention here that quantum breather excitations
have been seen in experiments as sharp discrete peaks in the
excitation spectrum of the ionic crystal NaI \cite{21}, as bound states of
up to seven phonons in PtCl \cite{22}, as two-phonon bound states in
the infrared absorption spectra of {\rm CO${}_{2}$} crystals \cite{23},
as a carrier of vibrational energy without dispersing for over more than
{\rm $10^{7}$} unit cells of layered crystal muscovite \cite{24}
etc. See \cite{12} for more details.

The plan of the paper is as follows: In Sec.\ 2 we describe the
Hamiltonian of the $\beta$-Fermi-Pasta-Ulam lattice and the
quantization scheme.  We write the Hamiltonian in terms of the phonon
creation and annihilation operators within the number conserving
approximation. In section 3 we describe the 
computational method. In section 4 we write the Hamiltonian under the mean field
approximation. In section 5 we write the complete and mean field Hamiltonians for NNN interaction
in terms of the phonon creation and annihilation operators and in section 6 we present the results and
discussion. Finally we conclude in section 7.

\section{The $\beta$-Fermi-Pasta-Ulam model and its quantization}
The Hamiltonian for the classical one dimensional $\beta$-FPU model for $N$ identical oscillators in a periodic lattice is
given by \cite{13}
\begin{equation}
  H = \sum_{j}\left[\frac{p_{j}^{2}}{2m} +\frac{\alpha}{2} 
    \left( x_{j}-x_{j-1}\right)^{2}
    +\frac{\beta}{4}\left( x_{j}-x_{j-1}\right)^{4}\right]
\end{equation}
where $\alpha$ is the nearest-neighbor harmonic force constant and
$\beta$ is the anharmonic force constant. The position and momentum
operators of the $j$-th particle can be written in terms of the phonon
creation and annihilation operators as
\begin{eqnarray*}
x_{j}&=&\sqrt{\frac{\hbar}{2m\omega}}\left(a_{j}^{\dagger}+a_{j} \right)\\
p_{j}&=&i\sqrt{\frac{\hbar}{2m\omega}}\left(a_{j}^{\dagger}-a_{j} \right)\\
\end{eqnarray*}
The creation and annihilation operators follow the commutation relations given by  
$[a_{j},a_{k}^{\dagger}]=\delta_{jk}$, and $[a_{j},a_{k}]=[a_{j}^{\dagger},a_{k}^{\dagger}]=0$.
Retaining only the number conserving terms, by virtue of the
lattice periodicity and assuming $N\rightarrow\infty$,where $N$ is the number of lattice sites, the Hamiltonian can be written as \cite{13}
\begin{eqnarray}
  H &=&\epsilon_{o}+\hbar\bar{\omega}\sum_{j} a_{j}^{\dagger}a_{j} 
-J'\sum_{j}a_{j}^{\dagger}\left( a_{j+1}+a_{j-1}\right) \nonumber \\
  & &+B\sum_{j}\left[a_{j}^{\dagger 2}a_{j}^{2}
    +\frac{1}{2}a_{j}^{\dagger 2}
\left( a_{j+1}^{2}+a_{j-1}^{2}\right) \right] \nonumber \\
  & &+B  \sum_{j}\sum_{s=\pm 1}\left[a_{j}^{\dagger}a_{j}a_{j+s}^{\dagger}a_{j+s} 
- \left( a_{j}^{\dagger 2}a_{j}a_{j+s} +{\rm h.c.}\right)\right]
\end{eqnarray}
where
\begin{eqnarray}
\epsilon_0&=&\frac{\hbar \omega N}{2}\left(1+\frac{3\beta}{2\alpha}
\left(\frac{\hbar}{2M \omega}\right)\right),  \\
&&\nonumber \\
\bar{\omega}&=&\omega\left(1+\frac{3\beta}{\alpha}
\left(\frac{\hbar}{2M\omega}\right)\right), 
\end{eqnarray}
\begin{eqnarray}
J'&=&\alpha\left(\frac{\hbar}{2 M \omega}\right)
\left(1+\frac{3\beta}{\alpha}\frac{\hbar}{M \omega}\right),  \\
&& \nonumber\\
B&=&3\beta\left(\frac{\hbar}{2 M \omega}\right)^{2}
\end{eqnarray}

\section{The Computational Method}
In Eqs. (1) and (2) we assume that the number of lattice sites is infinite, but in practice we can work with a finite number of sites. To simulate the infinite lattice, with finite number of sites, we incorporate the periodic boundary conditions (ring geometry). We consider a system with $f$ lattice sites.
Since we use the number conserving approximation, the Hamiltonian operator
commutes with the number operator
$\hat{N}=\sum_{j=1}^{f}a_{j}^{\dagger}a_{j}\;$, where $j=1,2,3,\ldots,f\;$
are lattice points. This implies that the operators $H$ and $\hat{N}$ have simultaneous eigenstates. Hence we can block-diagonalize the Hamiltonian
matrix using simultaneous eigenstates of $H$ and $\hat{N}$ as
\begin{equation}
H=\left(\begin{array}{cccc}
H_{0} & 0 & 0 & \cdots \\ 0 & H_{1} & 0 & \cdots \\ 
0 & 0 & H_{2} &\cdots \\
\vdots & \vdots & \vdots & \ddots \\
\end{array}\right)
\end{equation}
where each block $H_{m}\;$ describes states with m bosons. This greatly simplifies the calculations. Because of this block diagonal structure, it is possible to deal with each block independently \cite{17,18}.  For example in the paper we have calculated the energy spectrum for two quanta sector and four quanta sector. This corresponds to the blocks $H_{2}$ (Section 6.1) and $H_{4}$ (Section 6.2) respectively.

In the number state representation, the basis state is denoted by
$\left | \psi_n \right \rangle = [n_1, n_2, ....n_f]$, where $n_j$
denotes the number of quanta (bosons) at lattice site $j$.  For example
$\left[1\: 1\: 0\: 0\right]$, represents a two quanta state with four
lattice sites, with one boson each at the first and the other at the
second lattice sites and no bosons at the other two lattice sites, $N =
\sum_jn_j$.  For $m$ quanta and $f$ sites there are $\frac{(m+f-1)!}{m!(f-1)!}$ possible states  \cite{17,18}. For example for a 1D periodic lattice of $f=3$ sites and $m=2$ bosons, there are 6 possible states  $\left[2\: 0\: 0\: \right]$, $\left[0\: 2\: 0\: \right]$, $\left[0\: 0\: 2\: \right]$, $\left[1\: 1\: 0\: \right]$, $\left[0\: 1\: 1\: \right]$, and $\left[1\: 0\: 1\: \right]$ and therefore $H_2$ for this particular case is $6 \times 6$ matrix. 
  
To generate the Hamiltonian matrix, we make use of a hash table
\cite{25}.  To generate the label for each state, we assign a weight to
each lattice point as $w_{j}=\sqrt{p_{j}}\:$ where $p_{j}\:$ is the
$j^{th}\:$ prime.  The label for a state is given by,
$\sigma\left(\left[state\right]\right)=\sum_{j=1}^{j=f}w_{j}
n_{j}$, where $n_{j}$ is the number of quanta at the $j^{th}\:$
lattice point.  Use of the hash table considerably reduces the number
of operations needed to generate the Hamiltonian matrix.

Further, under the periodic boundary conditions, the Hamiltonian is
invariant under the action of the translation operator
$\hat{T}$. Hence we can use the translation invariant states to
further block-diagonalize each block of the Hamiltonian matrix \cite{17,18}. 
The number of translation invariant states ($\mathcal{N}$) for the system is
given by \cite{17,18}
\[\mathcal{N}=\frac{\left(m+f-1\right)!}{m!f!}
\]
Hence for 2 quanta and 3 sites, the number of translation invariant states is 2 for each value of $k$.
The two translation invariant states of a system with two
quanta and three lattice sites are given by
\begin{eqnarray}
\left|\psi_{1}\right\rangle & = & \left[2\: 0\: 0\right]
+t\left[0\: 2\: 0\right]+t^{2}\left[0\: 0\: 2\right]\\
\left|\psi_{2}\right\rangle & = & \left[1\: 1\:0\right]
+t\left[0\: 1\: 1\right]+t^{2}\left[1\: 0\: 1\right]
\end{eqnarray} 
where t=$e^{ik}$, with corresponding $k$ values $0,\pm2\pi/3$, so that
$t^{3}=1$. Thus we can further block-diagonalize $H_{2}$ into three $2 \times 2$ blocks $H_{2,k}$ using the translationally invariant states. The eigenvalues and eigenvectors of each of the $2 \times 2$ blocks can be calculated separately. This greatly reduces the memory requirement of each calculation. The allowed values of $k$ for a lattice with $f$ sites are $k = \frac{2\pi}{f}\nu$, where $\nu = 0, \ \pm 1, \ \pm 2, ... \pm(\frac{f}{2} -1), \ \pm \frac{f}{2}$ for $f$ even and $\nu = 0, \ \pm 1, \ \pm 2, ... \pm \frac{f-1}{2}$ for $f$ odd \cite{17}.
 
The number of invariant states increase rapidly with
increase in the number of quanta and also with the number of lattice
points. An arbitrary state vector of the system is given by
\begin{equation}
 \left | \Psi \right \rangle = \sum_j C_j\left | \psi_j \right \rangle 
\end{equation}
The normalization condition $\left \langle \Psi  | \Psi \right \rangle = 1$ gives $\sum_j \left | C_j \right |^2 = 1$.

Exact numerical diagonalization, using the LAPACK package \cite{26},
is then used to calculate the eigenvalues and eigenvectors of the
Hamiltonian matrix for the $m$-quanta sector on a lattice with $f$
sites.  To plot the energy spectrum we have considered
$\epsilon\left(k\right)=\left(E_{j}\left(k\right)-2\hbar
  \omega\right)/2J'$, where $E_{j}\left(k\right)$ are the eigenvalues
of the Hamiltonian matrix. The $k$ dependence of the energy $E_{j}(k)=\langle  \Psi| H | \Psi \rangle$ comes from the $t=e^{ik}$ dependence of the general state vector $| \Psi \rangle$ of the system (Eq.(10)), , with $k=\frac{2\pi \nu}{f}$, where $\nu=0,\pm 1, \pm 2, \ldots, \pm(f-1)/2$, through the translation invariant states $|\psi_{1}\rangle$ and  $|\psi_{2}\rangle$ (Eqs.(8 and 9)). \\
\noindent
{\it Spatial correlation function}: To identify breathers in the
eigenspectra so obtained, we calculate the spatial correlation
function to establish the main characteristic of spatial
localization. The spatial correlation of the displacements at sites
separated by $n$-lattice spacings is given by the expression \cite{20}
\begin{equation}
f_{\alpha}\left(j-j'\right)=\left\langle\alpha\right|\hat{x_{j'}}
\:\hat{x_{j}}\left|\alpha\right\rangle 
\end{equation}
where $\left|\alpha\right\rangle \:$ denotes the eigenvector of the
Hamiltonian and $x_{j}\:$ is the displacement at the $j^{th}\:$
site. Using the phonon creation and annihilation operators and the
number conserving approximation, we can write the above correlation
function as
\begin{equation}
 f_{\alpha}\left( j-j'\right)=\frac{\hbar}{2 m \omega}
\left\langle\alpha\right|\left(
  a_{j'}^{\dagger}a_{j}+a_{j'}a_{j}^{\dagger}
\right)\left|\alpha\right\rangle
\end{equation}

\section{ Mean field approximation :}

The last two terms in Eq. (2) describes bivibron tunneling \cite{13}. These 
terms  may be treated within the mean field approximation as an effective single vibron intersite tunneling and written as $2\langle a^{\dagger} a\rangle\sum_ja_{j}^{\dagger}(a_{j + 1} + a_{j - 1})$, where $\langle a^{\dagger} a\rangle$ denote the expectation value of the number operator $ a^{\dagger} a$ in an arbitrary state of the system. The details are given in the Appendix.
These two bivibron tunneling terms can now be added to the single vibron tunneling terms (3rd term in Eq. (2)) to get the mean field Hamiltonian with effective parameter $J=J'+2B\langle a^{\dagger} a \rangle\;$ \cite{13}. The effective Hamiltonian takes the form
\begin{eqnarray}
H_{{\rm MF}} &=&\epsilon_{o}+\hbar\bar{\omega}\sum_{j} a_{j}^{\dagger}a_{j}
-J\sum_{j}a_{j}^{\dagger}\left( a_{j+1}+a_{j-1}\right) \nonumber \\
& &+B\sum_{j}\left[a_{j}^{\dagger
    2}a_{j}^{2}+\frac{1}{2}a_{j}^{\dagger 2}
\left( a_{j+1}^{2}+a_{j-1}^{2}\right) \right] \nonumber \\
& &+B\sum_{j}\sum_{s=\pm 1}a_{j}^{\dagger}a_{j}a_{j+s}^{\dagger}a_{j+s} ,
\end{eqnarray}
Thus mean field approach approximates the direct bivibron tunneling process by effective single vibron tunneling. This allows approximate analytic solution of the problem as shown in \cite{13}.

\section{The Fermi-Pasta-Ulam lattice with nearest and 
next-nearest-neighbor interaction} 

Here we generalize the problem by including next-nearest neighbor
interactions to examine the effect of the long-range interaction on
the energy spectrum and correlation function of the system. Physical
systems such as polymers and biopolymers contain various charged
groups which interact with a long-range Coulomb force. Similarly, the
excitation transfer in molecular crystals and energy transport in
biopolymers are due to the transition dipole-dipole
interactions. Several studies of the classical discrete breathers in
presence of long-range interactions in one-dimensional chain have been
reported in the literature, for details see \cite{6}. Here we consider
the one-dimensional $\beta$- FPU chain with nearest and next-nearest
neighbor interactions. The classical Hamiltonian can be written as
\begin{eqnarray}
H &=& \sum_{j}\left[\frac{p_{j}^{2}}{2m}+\frac{\alpha}{2}
\left( x_{j}-x_{j-1}\right)^{2}+\frac{\alpha_{1}}{2}
\left( x_{j}-x_{j-2}\right)^{2} \right. \nonumber \\
& &\left. \qquad +\frac{\beta}{4}\left( x_{j}-x_{j-1}\right)^{4}
+\frac{\beta_{1}}{4}\left( x_{j}-x_{j-2}\right)^{4} \right]  \nonumber
\end{eqnarray}
Retaining only the number conserving terms and by virtue of the
lattice periodicity, the Hamiltonian can be written as
\begin{eqnarray}
H &=&\epsilon_{o}+2 \hbar\bar{\omega}\sum_{j} a_{j}^{\dagger}a_{j}
-J'\sum_{j}a_{j}^{\dagger}\left( a_{j+1}+a_{j-1}\right) \nonumber \\
& &+B\sum_{j}\left[a_{j}^{\dagger
    2}a_{j}^{2}+\tfrac{1}{2}a_{j}^{\dagger 2}
\left( a_{j+1}^{2}+a_{j-1}^{2}\right) \right] \nonumber \\ 
& &+B \sum_{j}\sum_{s=\pm 1}\left[ a_{j}^{\dagger}a_{j}a_{j+s}^{\dagger}a_{j+s}
-( a_{j}^{\dagger 2}a_{j}a_{j+s} +{\rm h.c.})\right] \nonumber  \\ 
& & -J_{1}'\sum_{j}a_{j}^{\dagger}\left(
  a_{j+2}+a_{j-2}\right) 
\nonumber  \\
& &+B_{1}\sum_{j}\left[a_{j}^{\dagger
    2}a_{j}^{2}+\tfrac{1}{2}a_{j}^{\dagger 2}
\left( a_{j+2}^{2}+a_{j-2}^{2}\right) \right] \nonumber \\
& &+B_{1}  \sum_{j}\sum_{s=\pm
  2}\left[a_{j}^{\dagger}a_{j}a_{j+s}^{\dagger}a_{j+s}
-( a_{j}^{\dagger 2}a_{j}a_{j+s}  +{\rm h.c.} )\right]
\end{eqnarray}
where the $J'$ and $B$ are given by Eqs.\ (5) and (6) respectively and
the $J_{1}'$ and $B_1$ are given by
\begin{eqnarray*}
J'_{1}&=&\alpha_{1}\left(\frac{\hbar}{2 M \omega}\right)
\left(1+\frac{3\beta_{1}}{\alpha_{1}}\frac{\hbar}
{M \omega}\right), \nonumber \\
&&\\
B_{1}&=&3\beta_{1}\left(\frac{\hbar}{2 M \omega}\right)^{2} \nonumber \\
\end{eqnarray*}
Using the mean field approximation as mentioned above, the mean field
Hamiltonian for the $\beta$-FPU lattice with NN and NNN interactions
in a given quanta sector can be written as
\begin{eqnarray}
H &=&\epsilon_{o}+2 \hbar\bar{\omega}\sum_{j} a_{j}^{\dagger}a_{j} 
\nonumber \\
& &-(J'+2 B \sum_{j} \left\langle a^{\dagger} a \right\rangle)
\sum_{j}a_{j}^{\dagger}
\left( a_{j+1}+a_{j-1}\right) \nonumber \\
& &+B\sum_{j}\left[a_{j}^{\dagger 2}a_{j}^{2}
+\tfrac{1}{2}a_{j}^{\dagger 2}\left( a_{j+1}^{2}+a_{j-1}^{2}\right) 
\right] \nonumber  
\end{eqnarray}
\begin{eqnarray}
& &+B \sum_{j}\sum_{s=\pm 1}a_{j}^{\dagger}a_{j}a_{j+s}^{\dagger}a_{j+s} \nonumber \\
& &-(J_{1}'+2 B_{1} \left\langle a^{\dagger} a 
\right\rangle )\sum_{j}a_{j}^{\dagger}\left( a_{j+2}+a_{j-2}\right) \nonumber  \\
& &+B_{1}\sum_{j}\left[a_{j}^{\dagger 2}a_{j}^{2}
+\tfrac{1}{2}a_{j}^{\dagger 2}\left( a_{j+2}^{2}+a_{j-2}^{2}\right) 
\right] \nonumber \\
& &+B_{1} \sum_{j}\sum_{s=\pm 2}a_{j}^{\dagger}a_{j}a_{j+s}^{\dagger}a_{j+s}
\end{eqnarray}

\section{Results and Discussions}
\subsection{\textbf{Two quanta sector}}
We first consider the discrete breather bands at the second quantum
level ($m=2$) or in the two-quanta sector. This corresponds to the block $H_{2}$ of the Hamiltonian matrix (Eq.(7)).  The $m=2$ state is obtained by operating two phonon creation operators on the ground state of the system. For example, for a lattice with three sites the $m=2$ state is given by the (Eq.(10)) along with (Eqns.(8 and 9)). For details see (Eq.(13)) of \cite{13}. We have considered a lattice of 101 lattice sites. As mentioned above, for odd number of sites the allowed maximum value of $k=\pm\frac{2\pi}{f}\frac{(f-1)}{2} \sim \pm \pi$ for large $f$. Therefore we plot the eigenspectrum for $-1\leq \frac{k}{\pi} \leq +1$. 

To begin with, we compare  our numerical exact diagonalization mean field result with the approximate analytical mean field result of Ivi\'{c} and Tsironis \cite{13} for the two-quanta sector. For easy reference, we have reproduced the mean field results of \cite{13} as Fig. 1(e). Fig. 1(a) shows our numerical results obtained for the choice of parameter value  $2B/J = -2$ which is  same as the parameter  value considered in \cite{13} (shown in Fig. \ 1.(e) by red lines).
Fig.\ 1.(e) shows that the two bands corresponding to the
on-site and nearest-neighbor (off-site) biphonon states (continuous and dotted red curves respectively) cross each
other at $k/\pi=\pm \frac{1}{2}$.  Ivi\'{c} and Tsironis attributed this
behavior to the approximate nature of their analytical calculations
and suggested to verify this fact through numerically exact
diagonalization method.  As mentioned in the introduction, one of the
motivations of our present work is to compare the analytic mean field
results of Ivi\'{c} and Tsironis with our exact diagonalization results.
From our study we find that that the crossings of the two bound
biphonon bands at $k/\pi=\pm \frac{1}{2}$ as obtained by Ivi\'{c} and Tsironis is indeed an artifact
of the approximate nature of their analytic calculations, as our
numerically exact diagonalization results (Fig. 1(a)) shows no such band crossings. Fig. 1(b) shows the energy spectrum for the complete Hamiltonian where the parameter $J^{'}$ is related to mean field parameters $J$ in Fig. 1(a) by the relation $J = J^{'} + 2B\langle a^{\dagger} a\rangle$. Fig. \ 1.(c) is the numerical mean field result for another set of parameter value $2B/J = -20$. This large value of the parameter ($2B/J$)  shows the separation of the two bound states very clearly (shown by the red lines). Again, our numerical mean field results do not show  any band crossing at $k/\pi=\pm \frac{1}{2}$. Fig. \ 1.(d) shows the eigenspectra for the complete Hamiltonian where the parameter $J^{'}$ is related to mean field parameters $J$ in Fig. 1(c) by the relation $J = J^{'} + 2B\langle a^{\dagger} a\rangle$. Fig.\ 1.(f) shows the results only for the complete Hamiltonian. This figure is plotted for large value of the parameter $|B| =1.2857$ to show larger band separation between the lowest two bands. The biphonon spectrum for repulsive interaction $B > 0$ is exactly symmetric to the one for
attractive interaction $B < 0$, but {\it above} the free two-phonon
band (not shown in the figure).
 However, contrary to the mean field case, the off-site
biphonon mode is not distinctly separated from the quasi-continuum
band even for large values of $2B/J'$.  This is in agreement with the results of
Xin-Guang and Yi \cite{14}.  To see why the off-site biphonon mode is
not distinctly separated from the quasi-continuum band, we consider
the matrix obtained for the two quanta mean-field Hamiltonian (Eq.\
(13)) given by
\begin{equation*}
H_{{\rm MF}}(k)=\left(
\begin{array}{c c c c c c c c}
D & E_{+} & 0   &0 &\cdots & \cdots  & 0 & 0 \\
E_{-} & H & F_{+}  & 0 & \cdots & \cdots & 0 & 0 \\
0 & F_{-} & 2   &  F_{+} & \cdots& \cdots & \vdots & 0 \\
\vdots & 0 & F_{-}  & \ddots & \cdots & \cdots & 0 \\
\vdots & \vdots & \vdots& \vdots   & \ddots & \cdots &\vdots & \vdots \\
\vdots & \vdots & \vdots& \vdots  & \vdots & \ddots & F_{+} & 0 \\
0 & 0 & 0    & \cdots & \cdots & F_{-} & 2 & F_{+} \\
0 & 0 & 0   & \cdots & \cdots & 0 & F_{-} & G
\end{array}
\right)
\end{equation*}
where $D(k) = 2 \hbar \omega + 2 B (1+ \cos(k))$, $E_{\pm}(k) = -
\sqrt{2}(1+\exp(\pm ik))J$, $F_{\pm}(k) = J (1+\exp(\pm ik))$, $G(k) = -2 J
\cos(\tfrac12(f+1) k)$ and $H = 2(1+B)$.  Similarly, the matrix
obtained for the complete Hamiltonian (Eq.\ (2)) is
\begin{equation*}
H_{comp}(k) = \left(
\begin{array}{c c c c c c c c}
D' & E'_{+} & 0   &0 & \cdots & \cdots  & 0 & 0 \\
E'_{-} & H' & F'_{+}  & 0 & \cdots& \cdots & 0 & 0 \\
0 & F'_{-} & 2   &  F'_{+} & \cdots &\cdots & \vdots & 0 \\
\vdots & 0 & F'_{-}  & \ddots &\cdots & \cdots & \cdots & 0 \\
\vdots & \vdots & \vdots &\vdots  & \ddots & \cdots &\vdots & \vdots \\
\vdots & \vdots & \vdots & \vdots & \vdots & \ddots & F'_{+} & 0 \\
0 & 0 & 0    & \cdots& \cdots &  F'_{-} & 2 & F'_{+} \\
0 & 0 & 0   & \cdots & \cdots& 0 & F'_{-} & G'
\end{array}
\right)
\end{equation*}
where $D'(k) = 2 \hbar \omega + 2 B (1+ \cos(k))$, $E'_{\pm }(k) = -
\sqrt{2}(1+\exp(\pm ik))(J'+B)$, $F'_{\pm}(k) = J' (1+\exp(\pm ik))$, $G'(k)
= -2 J' \cos(\tfrac12(f+1) k)$ and $H' = 2(1+B)$.  If we consider the
two matrices, we notice that the change in the nonlinearity parameter B
affects all the off-diagonal elements in $H_{MF}(k)$, whereas, it affects
only the first two off-diagonal elements of $H_{comp}(k)$. Hence the two
matrices are similar only for very small values of $B$. The presence
of the extra $B$ term in $H_{MF}(k)$ causes the lowering of the
eigenvalues corresponding to the off-site breather. Thus the off-site
breather appears separate from the continuum band in the mean field
case.
 
To show that the lowest isolated bands in Fig.\ 1 actually corresponds
to the quantum equivalent of the on-site classical discrete breather,
we establish their main characteristics, the spatial localization, by
calculating the spatial correlation function.

To calculate the spatial correlation function for the on-site discrete
breather, we consider the eigenvector $\left | \alpha \right \rangle$
(Eq.\ (12)) that corresponds to the lowest eigenvalue of the
Hamiltonian matrix (for complete Hamiltonian, Eq.\ (2)). Fig.\ 2(a) and
Fig.\ 2(b) show the spatial correlation function for the on-site
discrete breather for the centre and edge of the Brillouin zone
respectively. The parameters $B$ and $J'$ are the same as in Fig.\ 1.(f). From the figure we can see that the spatial correlation
function has a large peak at $n - n^{\prime} = 0$ and is almost zero
for other values of $n - n^{\prime}$. This means that the breather is
spatially localized at a single site $n$, which is a key characteristic
of onsite discrete breathers.  Fig.\ 2(c) and 2(d) shows the 
correlation functions for the mean field Hamiltonian (Eq.\ (13)). The values of the parameters $B$ and $J$ are same as the values in Fig. \ 1.(a). 

The localization property can also be seen from the plot of the
probability $\left |C_i \right |^2$ of the translational invariant
states $\left | \psi_i \right \rangle $ corresponding to the ground
state of the system \cite{18}. Fig.\ 3(a) and Fig.\ 3(b) show the
plots corresponding to the wavevector $k$ at the center and near the edge
of the Brillouin zone respectively. From the figures we notice that at
the centre of the Brillouin zone (Fig.\ 3(a)), the value of
$|C_{1}|^{2}$ is much greater than elsewhere. This implies that the
ground state system prefers to be in the on-site bound states. This
corresponds to the $2$-on-site breather band $[2,0,...]$ plus its
cyclic permutations (Eq.\ (8)). On the other hand, near the edge of
the Brillouin zone (Fig.\ 3(b)), the value of $|C_{2}|^{2}$ is much
greater than the rest. Hence the system prefers to be in the off-site
bound state $[1,1,0,...]$ plus cyclic permutations (Eq.\ (9)). These
results are in agreement with Xin-Guang and Yi \cite{14}. Fig.\ 3(c)
and Fig.\ 3(d) show the corresponding figures for the mean field
Hamiltonian (Eq.\ (13)) and it shows behavior similar to that for the
full Hamiltonian case (Eq.\ (2)).

\subsection{\textbf{Four quanta sector}}
So far we have restricted our study to the two-quanta
sector. Proceeding in the same way we have generalized the studies to
the case of the four-quanta sector, corresponding to the block 
$H_{4}$ of the Hamiltonian matrix. In this case we expect five bands 
as there are five possibilities of distributing four quantas on $f$ 
lattice sites. These are $[4,0,0,0,....], \ [3,1,0,0,....], 
\ [2,2,0,0,....], \ [2,1,1,0,....]$ and $[1,1,1,1,0,0....]$ and 
their cyclic permutations. Fig.\ 4 shows the eigenspectra for
the complete Hamiltonian (Eq.\ (2)) for four quanta sector for the
attractive interaction $B < 0$ case. Here we have considered 51
lattice sites. From the figure we see that the energy spectrum has the
same qualitative behavior as the two-quanta case (Fig.\ 1(d)) as
discussed above. However, here the continuum band splits into other
bands. As expected, we can observe 5 bands clearly. The overlap of the
3rd and 4th band as seen in the figure reduces with increase in the
strength of nonlinearity parameter $B$ and these two bands gets
separated. We have not shown the separated bands here, since for
larger value of $B$, the lowest band splits further away from the
other bands and it is difficult to show all the bands in the same
figure. The lowest band which is separated by a large magnitude
corresponds to the single $4$-on-site breather band, $[4,0,...]$ plus
cyclic permutations (similar to Eq.\ (8) for the two-quanta
case). The next lowest band corresponds to the 3-on-site breather band
plus single boson band $[3,1,0,...]$ plus cyclic permutations. The
third band is the double 2-on-site breather band $[2,2,0,...]$ plus
cyclic permutations, the 4-th band is the 2-on-site breather band plus
two single bosons $[2,1,1,0,...]$ plus cyclic permutations and 5-th
band (top band) consists of only single bosons $[1,1,1,1,0,...]$. It
is interesting to mention here that these five bands for the 4-quanta
sector as mentioned above also occur in the discrete nonlinear
Schrodinger (DNLS) model \cite{27}.

Fig.\ 5(a) shows the correlation function for the lowest band (Fig.\
4) of the four-quanta sector for 11 lattice sites.  From the figure we
can see that, as expected for the lowest band at the center of the
Brillouin zone ($k/\pi=0$), the correlation function have a large peak at
$n-n'=0$ and is almost zero for other values of $n-n'$.  The large peak
corresponds to 4-on-site discrete breather state. Small finite values
of the correlation function for non-zero values of $n-n'$ as seen in
the figure are due to finite size effect of the lattice.  This can be
seen from Fig.\ 5(b) where we have plotted the same correlation
function for 51 lattice sites and we can see that most of these peaks
at non-zero value of $n-n'$ disappear. Fig.\ 5(c) shows the
correlation function for the next lowest band (Fig.\ 4) of the
four-quanta sector for 11 lattice sites. As mentioned above, this band
corresponds to the 3-on-site breather band plus single boson band
$[3,1,0,...]$ plus cyclic permutations. From the figure we can see
that there is a large peak at $n-n'=0$ and oscillations of the
correlation function which have significant weight even at large
values of $n-n'$. The large peak corresponds to 3-on-site discrete
breather state and the oscillation corresponds to the one free-phonon
state. To show that the oscillation is not a finite size effect, we
have calculated the same correlation function for 51 lattice sites as
shown in Fig.\ 5(d). From the figure we can see that the oscillations
have significant weight even at $n-n'=6$.
\subsection{\textbf{Two quanta sector with next-nearest neighbour interaction}}
We now present the results for the one-dimensional $\beta$-FPU lattice
with nearest and next nearest neighbor interactions (Eq.\ (14)). This
is topologically equivalent to a zig-zag chain. If the lattice
periodicity for the straight chain is $a$, then for the zig-zag chain
it is doubled to $2a$. Accordingly the Brillouin zone boundary for the
zig-zag chain is at half its value ($\pi/2a$) as compared to that of
the straight chain. Thus the energy spectrum for the
one-dimensional $\beta$-FPU lattice with NN and NNN interactions is
expected to be qualitatively similar to that of the same lattice
system with only nearest neighbor interactions (Eq.\ (2)), except that
the Brillouin zone boundary of the NNN will be at half of the value
as that of the NN. This is exactly shown in Fig.\ 6 for the mean
field Hamiltonian (Eq.\ (15)) and in Fig.\ 7 for the full Hamiltonian
(Eq.\ (14)) for the two-quanta sector. Comparison of Fig.\ 6 with Fig
1(c) for the mean field case, we see that the nature of the energy
spectrum is similar in both cases, and the Brillouin zone boundary
in Fig.\ 6 occurs  at exactly half the value of the zone boundary in
Fig.\ 1(c). Comparison of Fig.\ 7 and Fig.\ 1(d) also shows similar
behavior for the respective full Hamiltonian cases. Fig.\ 8 (a) and
(b) show the plots for the correlation functions calculated for the
lowest mode of the eigenvalue spectrum (Fig.\ 7) at the centre and the
edge of the Brillouin zone respectively. This mode corresponds to on-site discrete
breather or the on-site bound biphonon state. The peak of the correlation
function at a particular site $n-n'=0$ shows the localization property
of the discrete breather. Fig.\ 8(c) and Fig.\ 8(d) are the plots of
the probability amplitudes at the center and edge of the Brillouin
zone respectively. From Fig.\ 8(c) we can see that there is a large
amplitude of $|C_1|^2$, which corresponds to the on-site discrete
breather or the on-site bound biphonon mode for $k=0$. On the other
hand at the edge of the Brillouin zone $|C_2|^2$ shows that the
2-off-site bound biphonon state is more probable. Similar behavior is
observed for the mean field Hamiltonian (Eq.\ (15)).

\section{Conclusion}
In conclusion, we have studied the quantization of the $\beta$-FPU
lattice with nearest and next-nearest neighbor interactions using
boson quantization rules and number conserving approximations. The
eigenvalue spectra in the two- and four-quanta sector are obtained
using an numerically exact diagonalization technique. From the calculation
of the energy spectrum and correlation function of the system, we have
shown that the quantum breathers exist as bound states of onsite and
nearest sites multi-phonons. We have shown that the probability
amplitude of the translational invariant states corresponding to a
quantum discrete breather state depends on the wave vector. For the
wavevector at the centre of the Brillouin zone, quantum breathers with
bound states of onsite multi-phonons are preferred. On the contrary,
for the wavevector at the edge of the Brillouin zone, the quantum
breathers with bound states of nearest sites multi-phonons are
preferred. The inclusion of next-nearest neighbour interactions do not
change the nature of the eigenspectra and the correlation functions of
the system except that the Brillouin zone is reduced to half. \\

\section{Appendix A:} 
The bivibron tunneling term $\sum_j(a_j^{\dagger})^2a_ja_{j\pm 1}$ can be written as:
$\sum_j(a_j^{\dagger})^2a_ja_{j\pm 1} = \sum_ja_j^{\dagger}a_j^{\dagger}a_ja_{j\pm 1} = \sum_j a_j^{\dagger}a_j^{\dagger}a_{j\pm 1}a_{j} = \sum_ja_j^{\dagger}a_{j\pm 1}a_j^{\dagger}a_j$
In the mean field approximation we take out the expectation value of the last two terms. This gives the  bivibron tunneling term as 
$\sum_j(a_j^{\dagger})^2a_ja_{j\pm 1} \approx \langle a^{\dagger}a\rangle(\sum_ja_j^{\dagger}a_{j\pm 1})$.

Similarly, the corresponding h.c. term can be written as:
$\sum_ja_{j\pm 1}^{\dagger}a_j^{\dagger}a_ja_j = \sum_j a_j^{\dagger}a_{j\pm 1}^{\dagger}a_ja_j = \sum_j a_j^{\dagger}a_ja_{j\pm 1}^{\dagger}a_j$. Under mean field approximation this term can be written as $ \approx \langle a^{\dagger}a\rangle\sum_ja_{j\pm 1}^{\dagger}a_j$ which for an infinite lattice ($N\rightarrow \infty)$ can be written as $\langle a^{\dagger}a\rangle \sum_ja_{j}^{\dagger} a_{j\mp 1}$.

We now show that $\langle a_j^{\dagger}a_j\rangle = \langle a^{\dagger}a\rangle = m$, independent of lattice site $j$ and depends only on the number of quanta $m$ in the state. For this we consider a system with three lattice sites and a state $|\Psi \rangle$ with two quanta ($m=2$). The state is given by Eq. (10) along with Eq. (8) and Eq. (9).  
\begin{eqnarray}
 a_1^{\dagger} a_1 |\Psi \rangle &=& C_1  a_1^{\dagger} a_1 |\psi_1 \rangle + C_2  a_1^{\dagger} a_1 |\psi_2 \rangle \nonumber \\
 &=& C_1\left ( 2[2 \ \ 0 \ \ 0] + 0 + 0  \right ) + C_2\left ( [1 \ \ 1 \ \ 0] +0 + t^2[1 \ \ 0 \ \ 1] \right )\nonumber 
 \end{eqnarray}
 Taking the scalar product with $\langle \Psi |$ and using orthonormality condition for the number states $[2 \ \ 0 \ \ 0], [0 \ \ 2 \ \ 0], .. [1 \ \ 1 \ \ 0]..$, we get 
 $$\langle \Psi |a_1^{\dagger} a_1 |\Psi \rangle = 2|C_1|^2 + |C_2|^2 + |C_2|^2 = 
2(|C_1|^2 +|C_2|^2) = 2$$
Similarly,
$$a_2^{\dagger} a_2 |\Psi \rangle = C_1\left ( 0 + 2t[0 \ \ 2 \ \ 0] + 0 \right ) + C_2\left ([1 \ \ 1 \ \ 0] + t[0 \ \ 1 \ \ 0] + 0 \right )$$
and
$$a_3^{\dagger} a_3 |\Psi \rangle = C_1\left ( 0 + 0 + 2t^2[0 \ \ 0 \ \ 2] \right ) + C_2\left (0 + t[0 \ \ 1 \ \ 1] + t^2[1 \ \ 0 \ \ 1] \right )$$
 Taking the scalar product with $\langle \Psi |$, we get
$$\langle \Psi |a_2^{\dagger} a_2 |\Psi \rangle = 2|C_1|^2 + |C_2|^2 + |C_2|^2 = 
2(|C_1|^2 +|C_2|^2) = 2$$
and,
$$\langle \Psi |a_3^{\dagger} a_3 |\Psi \rangle = 2|C_1|^2 + |C_2|^2 + |C_2|^2 = 
2(|C_1|^2 +|C_2|^2) = 2$$
This shows that $\langle a_{j}^{\dagger} a_{j} \rangle = \langle a^{\dagger} a \rangle $ is independent of $j$.
To show that $\langle a^{\dagger} a \rangle$ is independent of $f$, we consider a system with 5 lattice sites. For the two quanta sector $(m=2)$ and five lattice sites $(f=5)$ we get,
\begin{eqnarray*}
a_1^{\dagger} a_1 |\Psi \rangle &=& C_1 ( 2[2 \ \ 0 \ \ 0 \ \ 0 \ \ 0] + 0 + 0 + 0 + 0  ) + C_2 ([1 \ \ 1 \ \ 0 \ \ 0 \ \ 0]  + 0 \\
&+& 0 +0 + t^4[1 \ \ 0 \ \ 0 \ \ 0 \ \ 1]) 
+ C_3 ( [1 \ \ 0 \ \ 1 \ \ 0 \ \ 0] + 0 + 0 \\ &+& t^3[1 \ \ 0 \ \ 0 \ \ 1 \ \ 0] + 0  )
\end{eqnarray*}
 Hence,
 $$\langle \Psi |a_1^{\dagger} a_1 |\Psi \rangle = 2|C_1|^2 + |C_2|^2 + |C_2|^2 + |C_3|^2 + |C_3|^2 = 2(|C_1|^2 + |C_2|^2 + |C_3|^2) = 2$$
Similarly
$\langle\Psi | a_2^{\dagger}a_2|\Psi\rangle = \langle\Psi | a_3^{\dagger}a_3|\Psi\rangle = \langle\Psi | a_4^{\dagger}a_4|\Psi\rangle = \langle\Psi | a_5^{\dagger}a_5|\Psi\rangle = 2(|C_1|^2 + |C_2|^2 + |C_3|^2) = 2 $.
In general, for an arbitrary state $| \Psi \rangle$ with $m$ quanta and $f$ lattice sites $\langle \Psi | a_n^{\dagger}a_n | \Psi \rangle = m $ for all lattice sites i.e. $ n=1,2, \ldots, f$.\\

\noindent {\bf Acknowledgement:} BD would like to thank DST and BCUD-PU for
financial assistance through research projects and INSA-RSE for a visiting fellowship. BD would also like to
thank Heriot-Watt University, Edinburgh, UK and ICTP, Trieste, Italy
for hospitality, where part of the work was done.

\begin{figure}[H]
\begin{center}
\includegraphics[scale=0.9]{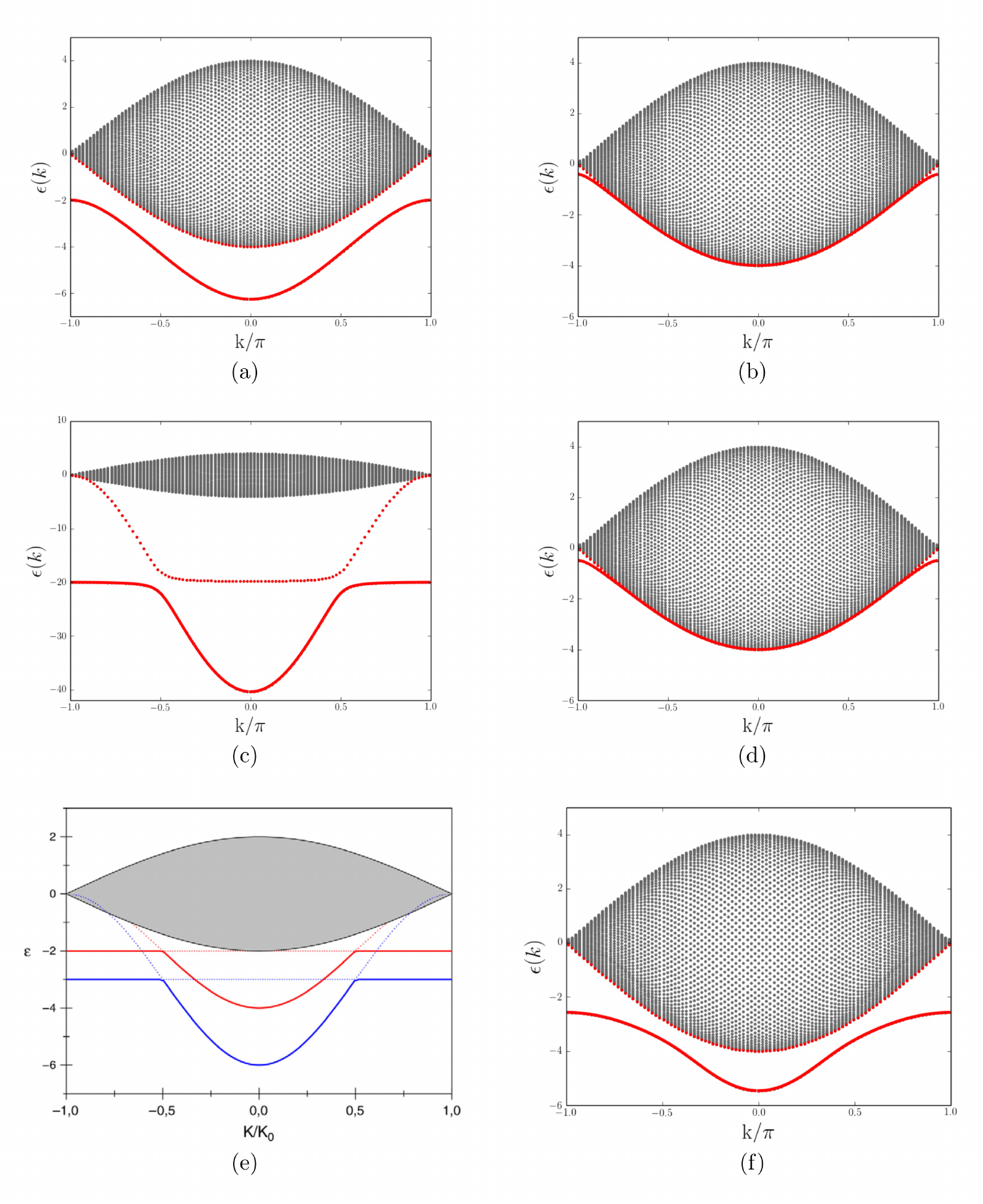}
\caption{Energy spectrum for(a) Mean field Hamiltonian for $B=-0.2$
  and $J=0.2$ (b) Complete Hamiltonian for $B=-0.2$ and $J'=1$ (c) Mean
  field Hamiltonian for $B=-0.2439$ and $J=0.02439$, (d) Complete Hamiltonian for
  $B=-0.2439$ and $J'=1$, (e) Mean Field Results of Ivi\'{c} and Tsironis, (figure obtained from \cite{13}) 
  (f) Complete Hamiltonian for $B=-1.2857$ and $J'=1$ }
\end{center}
\end{figure}

\begin{figure}[H]
\begin{center}
\includegraphics[scale=0.8]{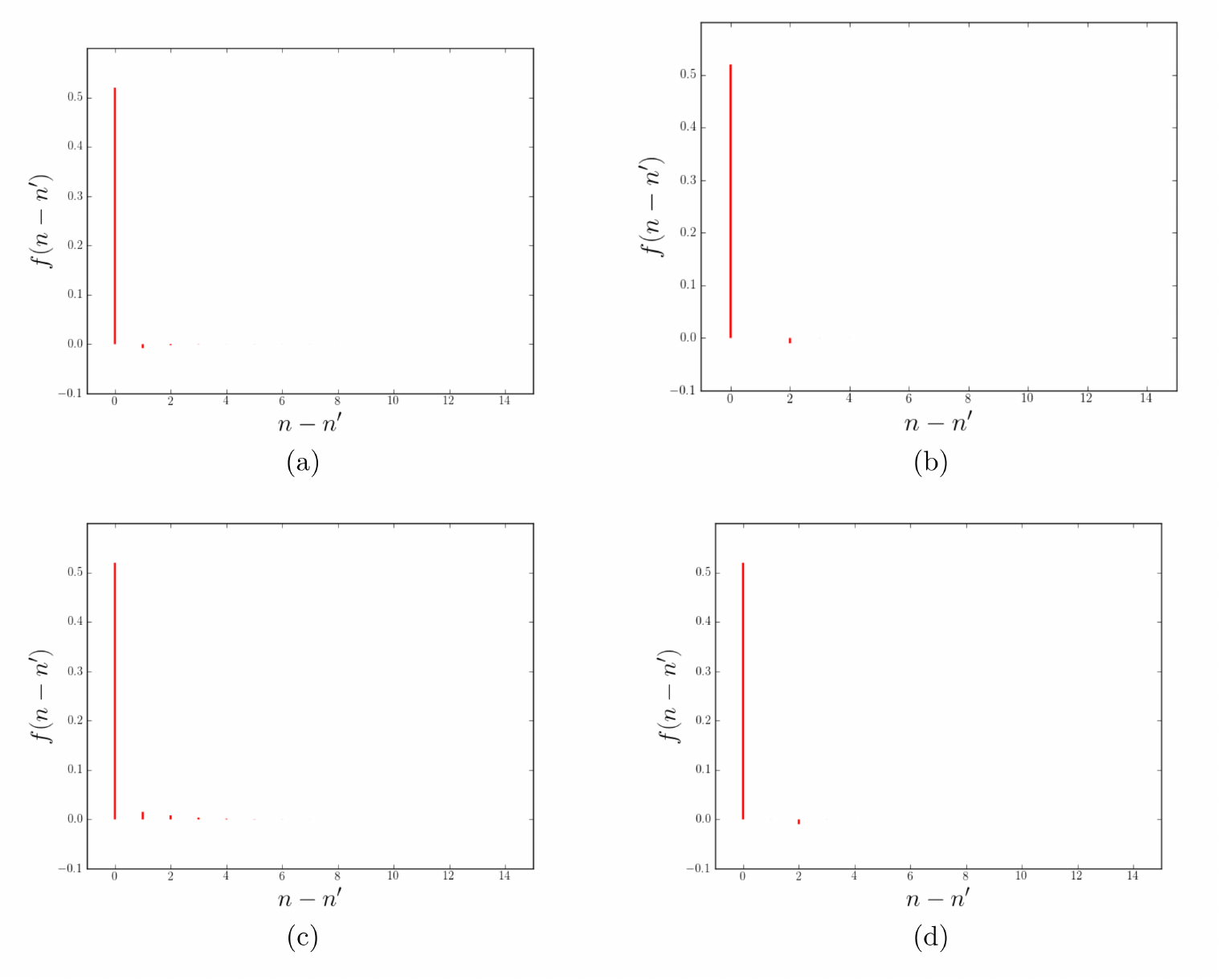}
\caption{(a)Spatial correlation
  Function for $k/\pi=0, B=-1.2857, J'=1$ for the complete Hamiltonian, (b)Spatial correlation Function for $k/\pi=50/51, B=-1.2857, J'=1$ for
  the complete Hamiltonian, (d) Correlation Function for $k/\pi=50/51,
  B=-0.2, J=0.2$ for the mean field Hamiltonian, (c) Correlation Function for $k/\pi=0, B=-0.2, J=0.2$ for
  the mean field Hamiltonian  }
\end{center}
\end{figure}
\begin{figure}[H]
\begin{center}
\includegraphics[scale=0.8]{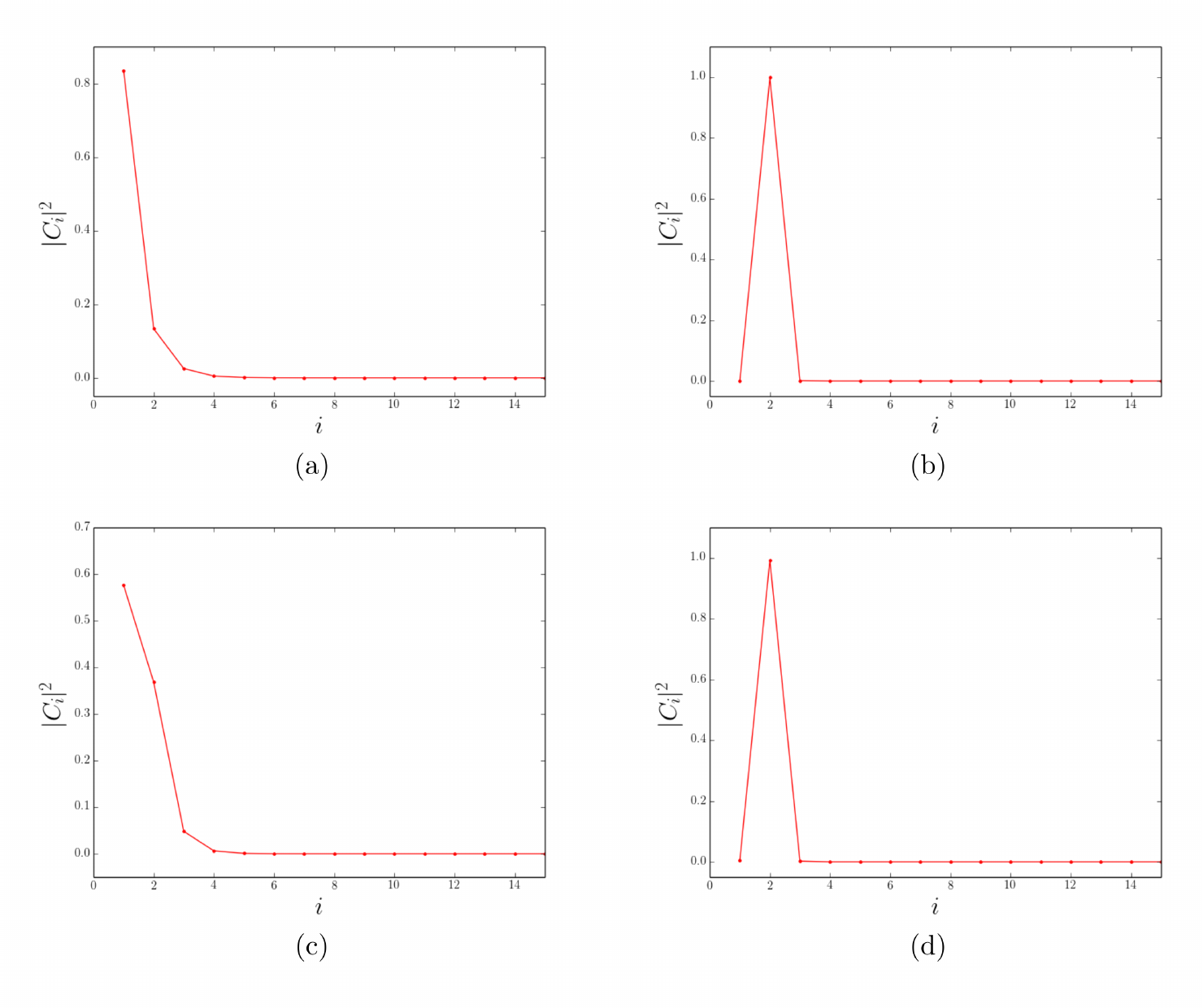}
\caption{ (a) $|C_{i}|^2$ for $k/\pi=0, B=-1.2857, J'=0.07$ for the complete
  Hamiltonian, (b) $|C_{i}|^2$ for $k/\pi=50/51, B=-1.2857, J'=0.07$ for the
  complete Hamiltonian, (c) $|C_{i}|^2$ for $k/\pi=0, B=-0.2, J=0.2$, for mean field Hamiltonian,
  (d) $|C_{i}|^2$ for $k/\pi=50/51, B=-0.2, J=0.2$, for mean field Hamiltonian }\end{center}
\end{figure}

\begin{figure}[H]
\begin{center}
\includegraphics[scale=0.7]{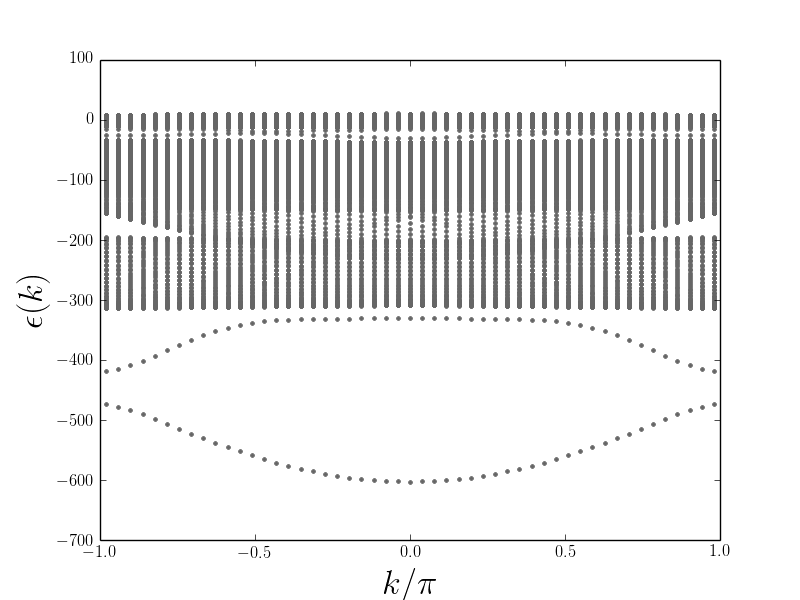}
\caption{Energy spectrum of the complete Hamiltonian for four quanta, $m = 4$,
  and 51 sites, $B=-20.0$ and $J'=1.0$}
\end{center}
\end{figure}

\begin{figure}[H]
\includegraphics[scale=0.9]{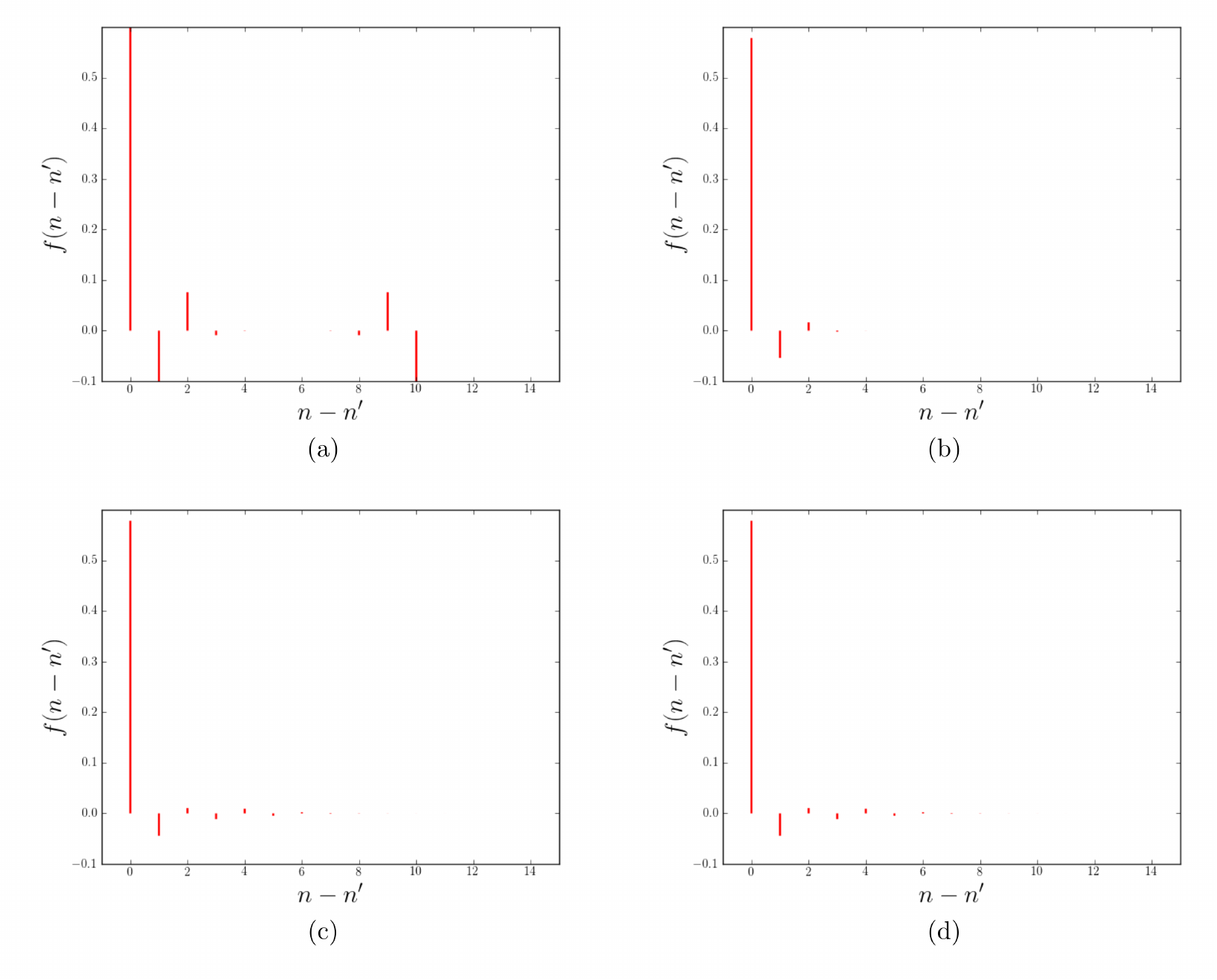}
\caption[]{ Spatial correlation function for the
  complete Hamiltonian, for four quanta, $m = 4$, with $k/\pi=0, B=-20.0$ and
  $J=1.0$, (a) for lowest energy mode with 11 sites, (b) for lowest
  energy mode with 51 sites, (c) for second lowest energy mode with 11
  sites, (d) for the second lowest energy mode with 51 sites. }
\end{figure}

\begin{figure}[H]
\begin{center}
\includegraphics[scale=0.7]{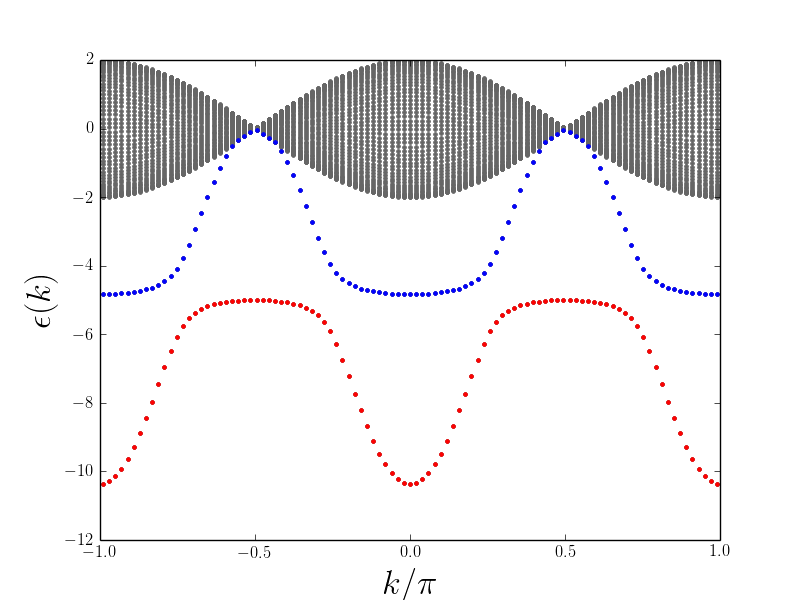}
\caption{Energy spectrum of the mean field Hamiltonian with nearest and next nearest neighbor interactions for, m=2,  $B=-5.0, J=1,
  B_{1}=-2.5, J_{1}=0.5$, 101 sites}
\end{center}
\end{figure}
\begin{figure}[H]
\begin{center}
\includegraphics[scale=0.7]{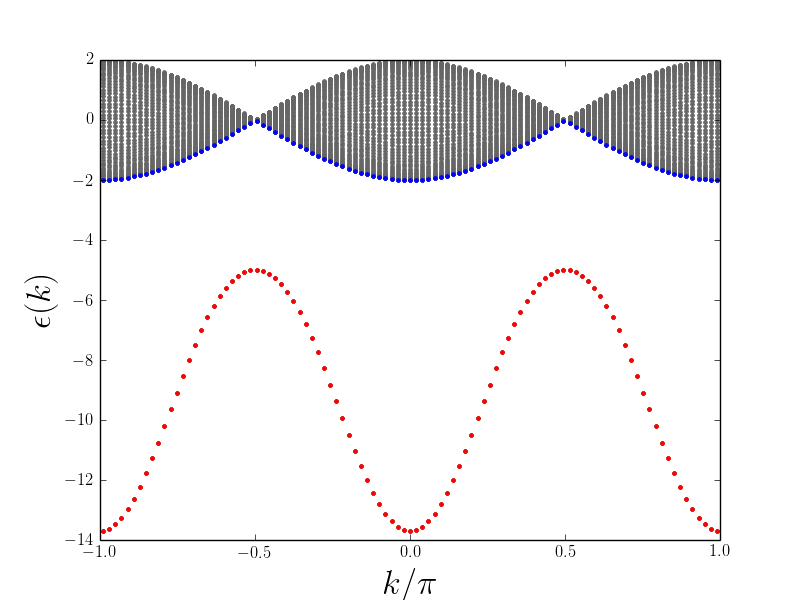}
\caption{Energy spectrum of the complete Hamiltonian with nearest and next nearest neighbor interactions for, m=2, $B=-5.0,
  J'=1, B_{1}=-2.5$, $J'_{1}=0.5$, 101 sites}
\end{center}
\end{figure}
\begin{figure}[H]
\begin{center}
\includegraphics[scale=0.8]{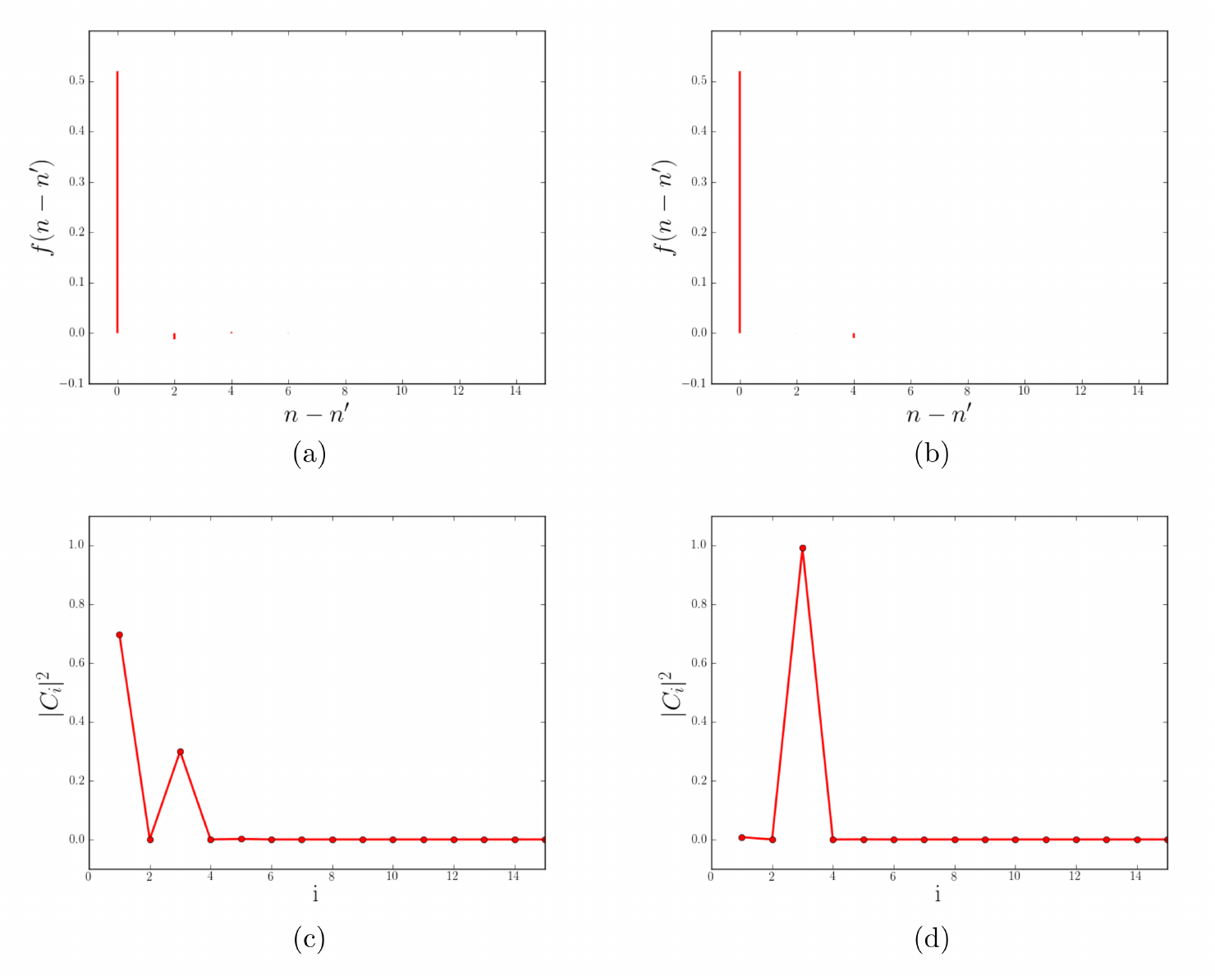}
\caption{(a) Spatial correlation function for $k/\pi=0, B=-5.0, J=1,
  B_{1}=-2.5$ and $J_{1}=0.5$, (b) Spatial correlation function for
  $k/\pi=50/51, B=-5.0, J=1, B_{1}=-2.5$ and $J_{1}=0.5$, (c) $|C_{i}|^2$
  for $k/\pi=0, B=-5.0, J=1, B_{1}=-2.5$ and $J_{1}=0.5$, (d) $|C_{i}|^2$
  for $k/\pi=50/51, B=-5.0, J=1, B_{1}=-2.5, J_{1}=0.5$ of the complete
  Hamiltonian with nearest and next nearest neighbor interactions}
\end{center}
\end{figure}
\end{document}